\documentclass[%
 reprint,
 amsmath,amssymb,
 aps,
 prl,
twocolumn,
superscriptaddress
]{revtex4}
\pdfoutput=1

\usepackage[dvipsnames]{xcolor}
\usepackage{graphicx}
\usepackage{dcolumn}
\usepackage{bm}
\usepackage[colorlinks=true,linkcolor=blue, citecolor=blue, urlcolor=blue, bookmarks]{hyperref}

\usepackage{ulem}

\begin{document}

\preprint{APS/123-QED}

\title{False vacuum decay in quantum spin chains}

\author{Gianluca Lagnese}
\affiliation{SISSA, via Bonomea 265, 34136 Trieste, Italy}
\affiliation{INFN, Sezione di Trieste, 34136 Trieste, Italy}
\author{Federica Maria Surace}
\affiliation{SISSA, via Bonomea 265, 34136 Trieste, Italy}
\affiliation{International Centre for Theoretical Physics (ICTP), Strada Costiera 11, 34151 Trieste, Italy}
\author{M\'arton Kormos}
\affiliation{MTA-BME Quantum-Dynamics and Correlations Research Group, E\"otv\"os Lor\'and Research Network (ELKH),
Budapest University of Technology and Economics, 1111 Budapest, Budafoki \'ut 8, Hungary}
\affiliation{BME ``Momentum'' Statistical Field Theory Research Group, Institute of Physics,
Budapest University of Technology and Economics, 1111 Budapest, Budafoki \'ut 8, Hungary}
\author{Pasquale Calabrese}
\affiliation{SISSA, via Bonomea 265, 34136 Trieste, Italy}
\affiliation{INFN, Sezione di Trieste, 34136 Trieste, Italy}
\affiliation{International Centre for Theoretical Physics (ICTP), Strada Costiera 11, 34151 Trieste, Italy}

\date{\today}

\begin{abstract}
The false vacuum decay has been a central theme in physics for half a century with applications to cosmology and to the theory of fundamental interactions.
This fascinating phenomenon is even more  intriguing when combined with the confinement of elementary particles.
Due to the astronomical time scales involved, the research has so far focused on theoretical aspects of this decay.
The purpose of this Letter is to show that the false vacuum decay is accessible to current optical experiments as quantum analog simulators of spin chains with confinement of 
the elementary excitations, which mimic the high energy phenomenology but in one spatial dimension.  
We study the non-equilibrium dynamics of the false vacuum in a quantum Ising chain and in an XXZ ladder. 
The false vacuum is the metastable state that arises in the ferromagnetic phase of the model when the symmetry is explicitly broken by a longitudinal field. 
This state decays through the formation of ``bubbles" of true vacuum. Using iTEBD simulations, we are able to study the real-time evolution in the thermodynamic limit and measure the decay rate of local observables. 
We find that the numerical results agree with the theoretical prediction that the decay rate is exponentially small in the inverse of the longitudinal field.
\end{abstract}

\maketitle

The possibility that our universe, as it cooled down, may have settled into a metastable
state (false vacuum) that may eventually decay was proposed by Coleman in 1977
and has been since then one of the most popularized ideas of physical cosmology~\cite{Coleman1977, Callan1977, Coleman1980, Turner1982}. The decay would happen through bubble nucleation, i.e. the formation of bubbles of true vacuum that rapidly expand: the probability for this process to occur is extremely small, and studying this phenomenon is notoriously challenging due to its intrinsic non-perturbative character.

Recently, the possibility of using tools from quantum technologies for studying problems of strongly coupled quantum field theories has attracted a lot of interest~\cite{Wiese2013,Dalmonte2016,Preskill2019,Banuls2020b}. On the one hand, tensor-network approaches are promising candidates for studying non-equilibrium properties that cannot be accessed with traditional Monte Carlo simulations. These approaches have been successfully applied to $1+1$ and $2+1$ dimensional lattice gauge theories~\cite{Byrnes2002,Banuls2013,Banuls2013b, Silvi2014,Tagliacozzo2014,Buyens2014,Rico2014,Banuls2020} but they suffer from limitations with dimensionality. Therefore, there has been an increasing interest in 
the toolbox of quantum simulators~\cite{Zohar2012,Banerjee2013,Mezzacapo2015,Martinez2016,Surace2020,Atas2021,Davoudi2021}. The hope is that controllable quantum systems in table-top experiments will help us understand difficult problems in quantum field theory, including, for example, the decay of the false vacuum~\cite{Billam2019,Billam2020,Abel2021,Ng2021,Billam2021}.

In this context, one-dimensional quantum spin models represent the ideal framework for benchmarking quantum simulators: they can host particle confinement~\cite{McCoy1978,Ishimura1980}, a property that can be observed in the non-equilibrium dynamics after a quantum quench~\cite{Kormos2016,Liu2018,tan2019}; 
it has also been suggested that their real-time evolution can reveal interesting phenomena including collisions of particles and bubble nucleation~\cite{Surace2021,Karpov2020,Milsted2020,Sinha2021,vcc-20,Rigobello2021}. 

Here, we propose to study the decay of the false vacuum in quantum spin models using simulations of real-time dynamics after a quantum quench. 

\begin{figure}
    \centering
    \includegraphics[width=\columnwidth]{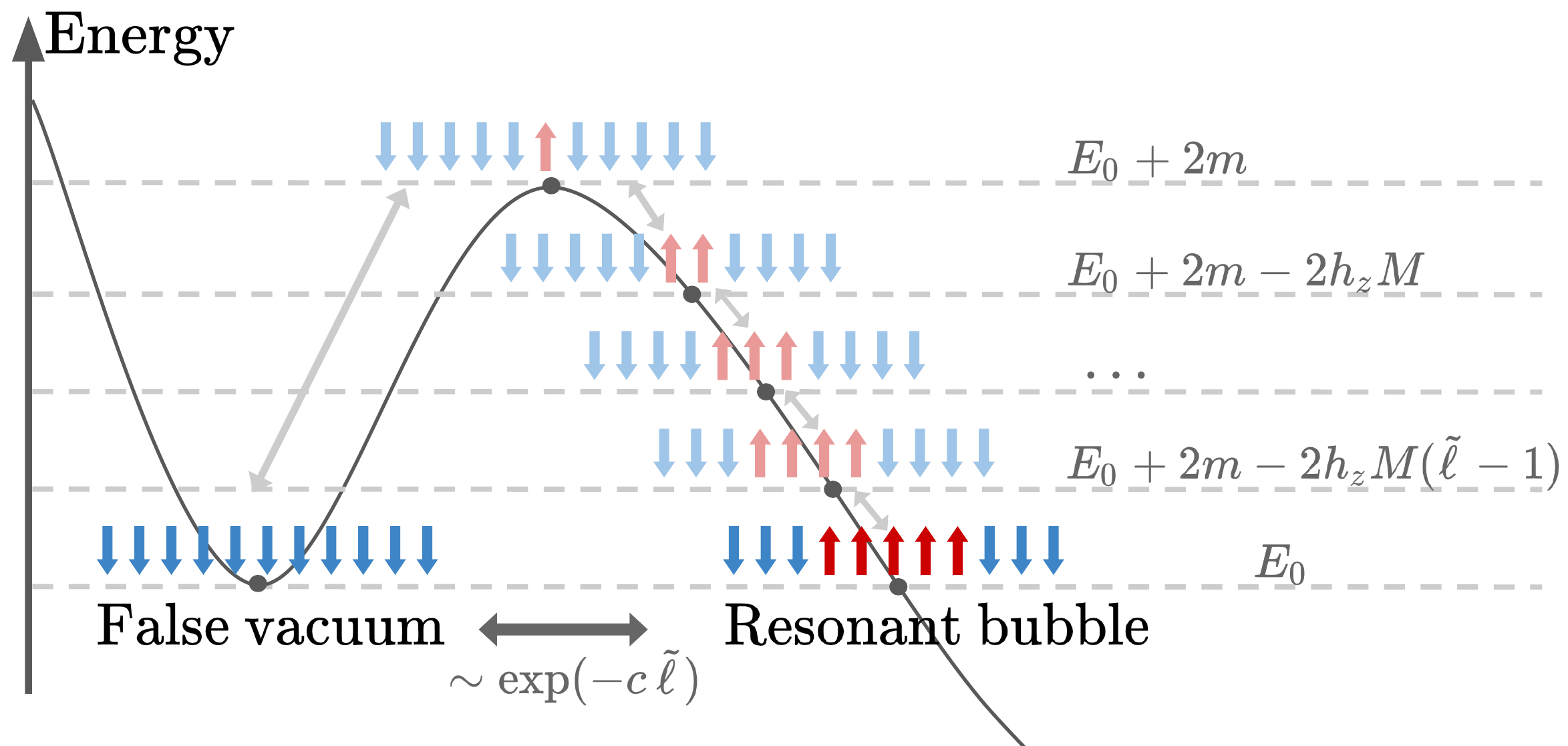}
    \caption{{\it Illustration of bubble formation} (false vacuum is in blue, true vacuum in red). The process that leads to the resonant bubble goes through $O(\tilde \ell)$ off-resonant states: a small bubble is (virtually) created and expanded until it reaches the resonant size $\tilde \ell$. As a consequence, the matrix element that drives the false vacuum decay is exponentially small in $\tilde \ell\propto h_z^{-1}$.}
    \label{fig:sketch}
\end{figure}

\paragraph{False vacuum decay---}
To illustrate the phenomenon of false vacuum decay we consider a prototypical model for confinement, the quantum Ising chain in transverse and longitudinal field, with Hamiltonian
\begin{equation}
\label{eq:IsingHam}
    H(h_x, h_z) = -\sum_i \left(\sigma_i^z \sigma_{i+1}^z+h_x \sigma_i^x+h_z \sigma_i^z\right)\,,
\end{equation}
where $\sigma_i^\alpha$ are Pauli operators, and the amplitudes $h_x$ and $h_z$ are the transverse and longitudinal field, respectively.

For $h_z=0$ the model has a $\mathbb{Z}_2$ symmetry that is spontaneously broken for $|h_x|<1$ (ferromagnetic phase). In this phase there are two ground states characterized by opposite magnetizations $\langle \sigma_i^z \rangle=\pm M$, with $M=(1-h_x^2)^{1/8}$ \cite{sach-book}. The model is diagonalized with a mapping to free fermions: the corresponding excitations in the ferromagnetic phase are kinks that interpolate between domains with opposite magnetization \cite{m-book}. 
The kinks can propagate freely and have dispersion relation
\begin{equation}
\label{eq:dispersionrelation}
    \omega(\theta) = 2\left(1-2h_x \cos \theta +h_x^2 \right)^{1/2}.
\end{equation}

For a longitudinal field $h_z \neq 0$, the $\mathbb{Z}_2$ symmetry is explicitly broken and the degeneracy between the two ground states is split by an extensive quantity $\sim 2h_z M N$, where $N$ is the number of sites in the chain: the state with magnetization aligned with the external field (the {\it true vacuum}) is the ground state of the model, while the one with opposite magnetization (the {\it false vacuum}) is a metastable state.
The nature of the excitations is also drastically modified: the longitudinal field induces a linear potential between the kinks, confining them into {\it mesons} \cite{McCoy1978}.
The confinement strongly affects many aspects of the non-equilibrium dynamics \cite{Kormos2016,Liu2018,tan2019,Surace2021,Karpov2020,vcc-20,Cai2012,jkr-19,rjk-19,mplcg-19,clsv-20,vk-20,mrw-17,cr-19,ylgi-20,pp-20,rmc-16}. 

The false vacuum is at high energy, so it can resonantly decay into the continuum of multi-meson states. While this decay is a very complicated process, the basic mechanism can be understood as the formation of bubbles of true vacuum in the system. Creating a bubble of size $\ell$  requires the energy given by the masses of the two kinks lowered by $2h_zM\ell$. When this energy becomes zero, the bubble is resonantly excited. This bubble can then further decay through other resonant processes. 
However, for $h_z$ sufficiently small, the phenomenon of bubble formation is very slow. 
This slowness can be understood by the following simple heuristic argument.  
A resonant bubble of size $\tilde \ell$ results from the frequent creation of a small bubble (of size of order 1) that 
then should expand until it reaches the resonant size $\tilde \ell \gg 1$ (see Fig.~\ref{fig:sketch}). 
This expansion is a high-order process in the perturbation theory in $h_z$ and, 
as a consequence, the matrix element for exciting the resonant bubble is exponentially small in $\tilde \ell \propto h_z^{-1}$.


 The decay of the metastable false vacuum in the Ising chain has been studied in Ref. \onlinecite{Rutkevich1999FVD}, where the following expression of the decay rate per site was obtained \footnote{In Ref. \onlinecite{Rutkevich1999FVD},  the rate $\gamma$ contains an oscillatory term $g(h_z)$: we work here in the approximation $g(h_z) \simeq 1$, which is justified  for $h_z$ sufficiently far from $1$.}:
\begin{equation}
\label{eq:gamma}
    \gamma = \frac{\pi}{9} h_z M \exp\left(-\frac{q}{h_z}\right)
\end{equation}
with $q = |f(-i\ln h_x)|/M$ and $f(\theta)=2\int_0^\theta \omega(\alpha)\, \mathrm{d}\alpha$. Note that $q$ and $M$ only depend on $h_x$. 
This rate $\gamma$ can be interpreted as the number of resonant bubbles that are created per unit time divided by the number of sites. In agreement with the argument explained above, the decay is non-perturbative in the longitudinal field, with an exponential dependence on $h_z^{-1}$. 

We note that an analogous mechanism drives the phenomenon known as {\it string breaking}. String breaking is typically understood as the saturation of the effective interaction between two static charges (or kinks, in this case) at large distance, due to the screening effects of other charges: in other words, the string that extends between the two static charges is broken by the creation of dynamical charges. In the model we are studying, the string corresponds to a false vacuum domain and the string breaking effect corresponds to the formation of a bubble in the domain. The dynamics of string breaking has been studied in this model, in other spin chains, and lattice gauge theories~\cite{Banerjee2012,Hebenstreit2013,Kuhn2015,Pichler2016,Kasper2016,Buyens2017,Kuno2017,Sala2018,Spitz2019,Park2019,Surace2020,Magnifico2020,Notarnicola2020,Chanda2020,Lerose2020,Verdel2020}, and similar expressions for the decay rate were found.

\paragraph{Quench protocol and methods---}

The goal of this Letter is to show that a window of Hamiltonian parameters of the Ising spin chain ($h_x,h_z$) exists such that the false vacuum decay 
can be observed through numerical simulations of the non-equilibrium dynamics after a quantum quench. 
The quench protocol is the following: 
i) we prepare the system in the ferromagnetic state with all the spins in the $\sigma_i^z=1$ direction; 
ii) we evolve the system in imaginary time with the Hamiltonian $H(h_x, -h_z)$ using infinite volume time evolving block decimation (iTEBD) until we achieve a good convergence to the ground state; 
iii) we quench $-h_z\rightarrow h_z$ and evolve in real time.
Using this protocol, we are able to prepare the false vacuum of $H(h_x, h_z)$ and study its evolution in real time using iTEBD.
The state preparation ii) is obtained using a Trotter step $\delta t=10^{-3}$, and the imaginary time evolution stops when the relative change of the energy density is smaller than $10^{-16}$.
The real time evolution after the quench iii) is performed with a Trotter step $\delta t=10^{-2}$. The bond dimension $\chi$ is set to $512$. We checked the stability of the numerical simulations with respect to changes in $\chi$ and $\delta t$.

We stress that in our quench protocol the false vacuum decay drives the system toward a thermal state, that has a finite energy density with respect to the true vacuum.
Only in the limit  $h_z\to 0$ this state tends to the true vacuum. 

\paragraph{Time scales---} Before embarking on the analysis of the numerical data, we should have a clear picture of all
the time scales entering in the quench dynamics of our model.
Starting from the false vacuum, the first process happening is the creation of off-resonant bubbles. 
During this (relatively) short-time transient, say up to time $\tau_\text{r}$, the system remains effectively frozen in the false vacuum until the resonant bubbles start being produced.
However, here we are not interested in this transient but only in 
the growth of the resonant bubbles, because this is the process that leads to the false vacuum decay described by the rate \eqref{eq:gamma}.
For the accurate measurement of this rate, we need a clear separation of this time scale from the successive ones.
%
%
Indeed, at very late time, when most of the false vacuum decayed, since the system is at finite energy density, it starts thermalizing 
through the propagating states that originate from the decay of the resonant bubbles:
the late time dynamics is governed by the thermal state corresponding to the energy of the pre-quench state (only for very small $h_z$ this is close to 
zero temperature, i.e. the true vacuum). We denote with $\tau_D$ the time scale for the onset of thermalization;
unfortunately, we do not know how to estimate $\tau_\text{D}$, but its determination lies beyond the scope of this Letter.

We emphasize that Eq.~\eqref{eq:gamma} is expected to work well under the assumption of a clean separation of time scales,  
i.e. $\tau_r\ll \gamma^{-1}\ll \tau_D$. 
For the Hamiltonian \eqref{eq:IsingHam}, such separation of time scales is guaranteed in the regime $h_z \ll 1$ and $h_x$ not too close to $1$. 
The requirement $h_z \ll 1$ is obvious, since as $h_z$ grows all the above time scales $\tau_r, \gamma^{-1}, \tau_D$ become of order one and there cannot be any separation. 
Moreover, if $h_x$ gets too close to $1$, the masses of the kinks become very small and the assumption that the false vacuum preferably decays into one-domain states
(resonant bubbles) is no longer justified.

In conclusion, the false vacuum decay is expected to be described by Eq. \eqref{eq:gamma} in the limit of small $h_z$ and $h_x$ not too close to $1$.
However, as the fields are reduced, the time scale $\gamma^{-1}$ soon becomes extremely large (which is the reason why false vacuum decay is generically an elusive 
phenomenon, see also \cite{pwzt-21}).
Thus the main difficulty of the numerical analysis is to find a window of the Hamiltonian parameters such that there is an optimal balance between 
a reasonable separation of times scales (to have a time range in which Eq. \eqref{eq:gamma} describes something) and its numerical 
accessibility. We found that such balance is obtained for rather small $h_z$ (of the order of $10^{-2}$), but with $h_x$ relatively large $h_x\sim[0.7,0.9]$:
a smaller $h_x$ makes the decay time ($\gamma^{-1}$) too long and a larger $h_z$ destroys completely the time-scale separation.

\paragraph{Results---}
To estimate the decay rate, we analyze the following two observables
\begin{align}
    F(t) &= \frac{\langle \sigma_i^z(t)\rangle+\langle \sigma_i^z(0)\rangle}{2 \langle \sigma_i^z(0) \rangle},
    \label{Fd}\\
    G(t)&=1-||\rho(t)- \rho(0)||_1,
    \label{Gd}
\end{align}
where $\rho(t)$ is the two-site density matrix at time $t$ and $||\rho(t)- \rho(0) ||_1$ is the trace distance between the two density matrices. 
Both quantities can be easily computed in iTEBD, and satisfy $F(0)=G(0)=1$, while they vanish in the true vacuum. 
The time evolution of $F(t)$ is fully encoded in the magnetization and, consequently, is expected to decay with a rate
\begin{equation}
\label{eq:gammaF}
    \gamma_F \simeq \gamma \tilde \ell= \frac{f(\pi)}{18} \exp \left(-\frac{q}{h_z}\right),
\end{equation}
where the size of the resonant bubble is $\tilde{\ell}=\frac{f(\pi)}{2 h_z M \pi}$ (see Ref. \onlinecite{Rutkevich1999FVD}).
Note that for small $h_z$, this rate is much larger than $\gamma$, so the time scale needed to observe the decay in our simulation is significantly reduced.

\begin{figure}[t]
    \centering
    \includegraphics[width=0.5\textwidth]{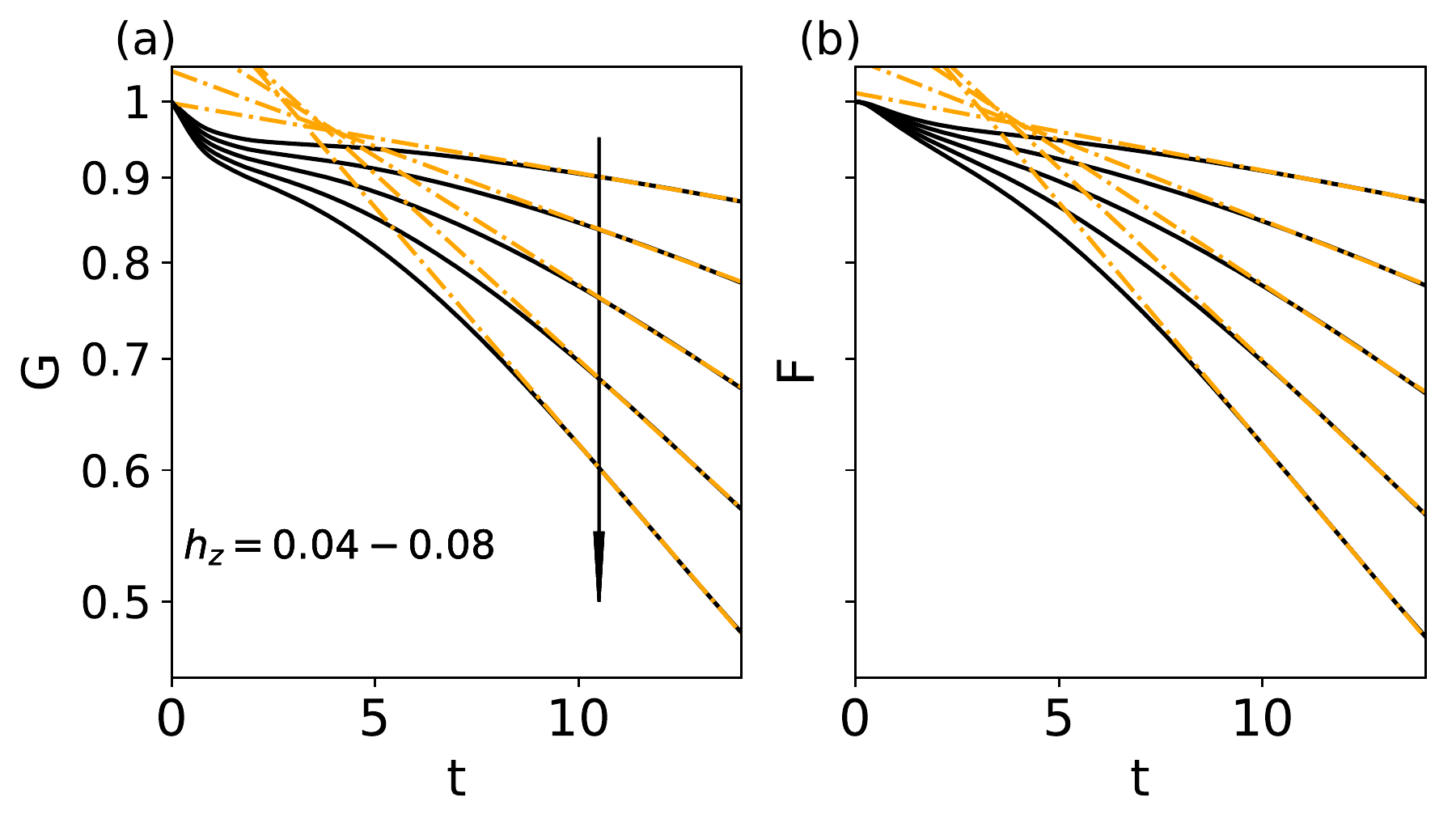}
    \caption{{\it False vacuum decay in the quantum Ising chain}. The time evolution of $F(t)$ and $G(t)$, in Eqs. \eqref{Fd} and \eqref{Gd}, is shown for $h_x= 0.8$ and different values of $h_z$ after the quench $-h_z\to h_z$. The dot-dashed lines are the exponential fits in the decay region performed to extract the decay rates $\gamma_{F}$ and $\gamma_{G}$. }
    \label{fig:ising1}
\end{figure}

As an illustrative example for the determination of the decay rates of $F$ and $G$, 
in Fig.~\ref{fig:ising1} we report their evolution at fixed $h_x=0.8$ and different values of $h_z$ on a semi-log scale. 
It is evident that after a short transient, all the data show a distinct exponential decay (linear behavior on semi-log scale). 

\begin{figure*}[t]
    \centering
    \includegraphics[width=\textwidth]{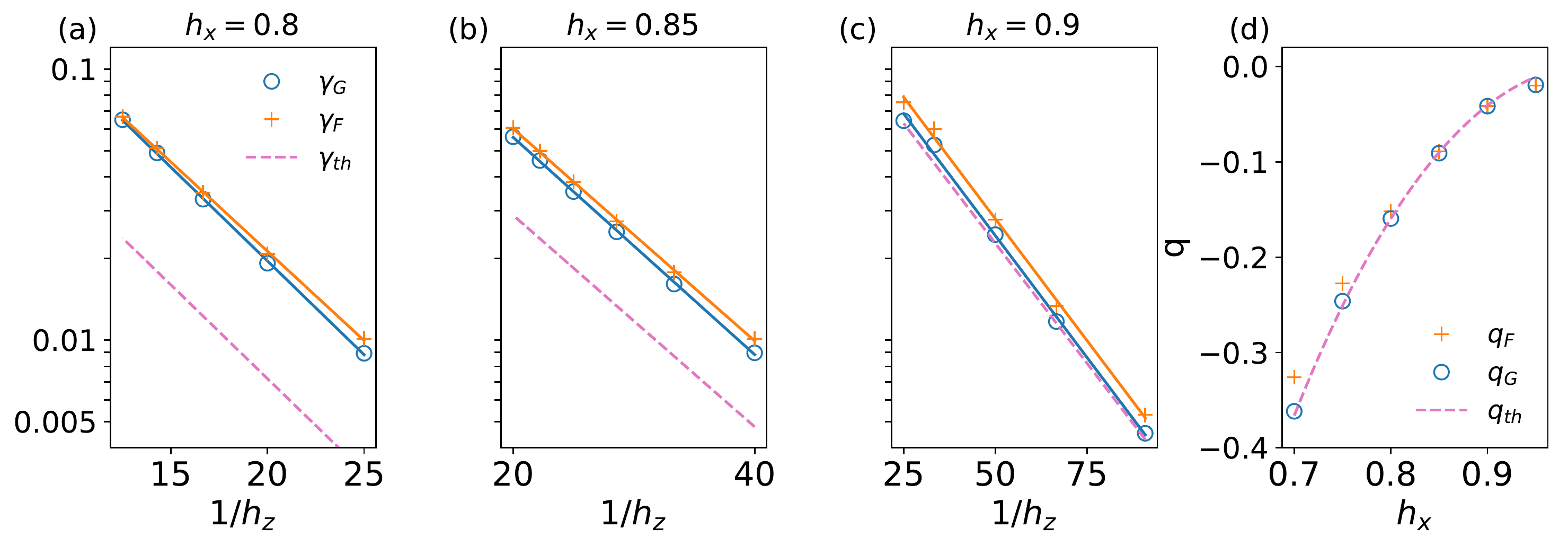}
    \caption{{\it Decay rates in the quantum Ising chain}. 
    In panels (a),(b),(c) we show the decay rates $\gamma_F, \gamma_G$, obtained from the fits of $F(t)$, $G(t)$ as in Fig.~\ref{fig:ising1}. 
    The continuous lines are the fits of the exponential dependence of the rates in $1/h_z$. 
    The dashed line represents the theoretical prediction~\eqref{eq:gammaF}. 
    From the fits the coefficients $q_F$ and $q_G$ are extrapolated and showed against
     the theoretical value $q_{\rm th} =|f(-i\ln h_x)|/M$ (dashed line) in panel (d). }
    \label{fig:ising2}
\end{figure*}

For all the considered values of $h_x$ and $h_z$, we performed an exponential fit $O(t)=A_O e^{-\gamma_O t}$, with $O=F,G$.
The fit is done in a time range $t_0<t<t_1$ and then we check the stability of the fit for small variations of $t_0,t_1$.
The resulting decay rates $\gamma_{F,G}$ are plotted in Fig. \ref{fig:ising2}-a,b,c as functions of $h_z^{-1}$ again on semi-log scale.
The exponential dependence on $1/h_z$, expected from Eq.~(\ref{eq:gammaF}), is very clear in the data. 
We fitted these rates with 
\begin{equation}
\label{eq:td0}
    \gamma_{O} = k_{O} e^{ -q_{O}/h_z}, \qquad O=F,G.
\end{equation}
In Fig. \ref{fig:ising2}-d we report the obtained coefficients $q_F$, $q_G$: they are compatible with each other and they both 
 agree very well with
the theoretical prediction $q = |f(-i\ln h_x)|/M$ in the full range of $h_x$ considered.
The prefactors $k_F$ and $k_G$ 
in Eq. \eqref{eq:td0} turn out to be different from what predicted by Eq. \eqref{eq:gammaF}
(the data in Fig. \ref{fig:ising2}-a,b,c are shifted compared to the dashed line).  
However,  this shift is not surprising at all because we know that (i) the prefactor depends on the specific observable
(e.g., compare Eqs. \eqref{eq:gamma} and \eqref{eq:gammaF}),
(ii) we expect it to be more affected by the approximations done in the derivation of Eq.~(\ref{eq:gammaF}).


\paragraph{XXZ ladder---}
To show the general validity of our analysis, we consider  a second model presenting confinement of elementary excitations with hamiltonian \cite{gian20}
\begin{equation}
\begin{split}
\label{eq:ladder}
	    H(\Delta_{||},\Delta_{\perp}) = &\frac{1}{2} \sum_{j=1}^{L}
	    \sum_{\alpha=1,2}\left[ \sigma^x_{j,\alpha} \sigma^{x}_{j+1,\alpha} +
	    \sigma^{y}_{j,\alpha} \sigma^{y}_{j+1,\alpha} \right.\\
	    +& \left. \Delta_{||}\sigma_{j,\alpha}^{z} \sigma_{j+1,\alpha}^{z}\right]+ \Delta_{\perp}\sum_{j=1}^{L} \sigma^{z}_{j,2} \sigma^{z}_{j,1}\,
\end{split}
\end{equation}
i.e. two XXZ spin-1/2 chains coupled along the longitudinal direction through an anisotropic Ising-like interaction.
Compared to the Ising spin chain~\eqref{eq:IsingHam}, the model possesses two interesting features. 
The first is that in the absence of the confining interaction ($h_z$ and $\Delta_\perp$), the Ising spin chain becomes a free model, while the decoupled XXZ chains constitute an interacting (integrable) spin model. 
The second one is that confinement is 
induced by the internal interaction between the chains, a built-in mechanism, instead of an external field (and this is more similar to what happens for quarks).
We work in the gapped anti-ferromagnetic phase, i.e. $\Delta_{||} \in (1, +\infty) $ where
the model for $\Delta_{\perp}=0$ has four degenerate antiferromagnetic ground states. 
The confining potential explicitly breaks the original $\mathbb{Z}_2 \times \mathbb{Z}_2$ symmetry to a single $\mathbb{Z}_2$ \cite{gian20}: 
the four degenerate ground states at $\Delta_{\perp}=0$ are split in two doublets separated by an energy of the order $\Delta_{\perp}L$. 
The two lowest states (the true vacua) are now the stable ground states, while the other two (the false vacua) are metastable states at high energy and can decay in 
the continuum of the many-body spectrum. 

\begin{figure}[b]
    \centering
    \includegraphics[width=\columnwidth]{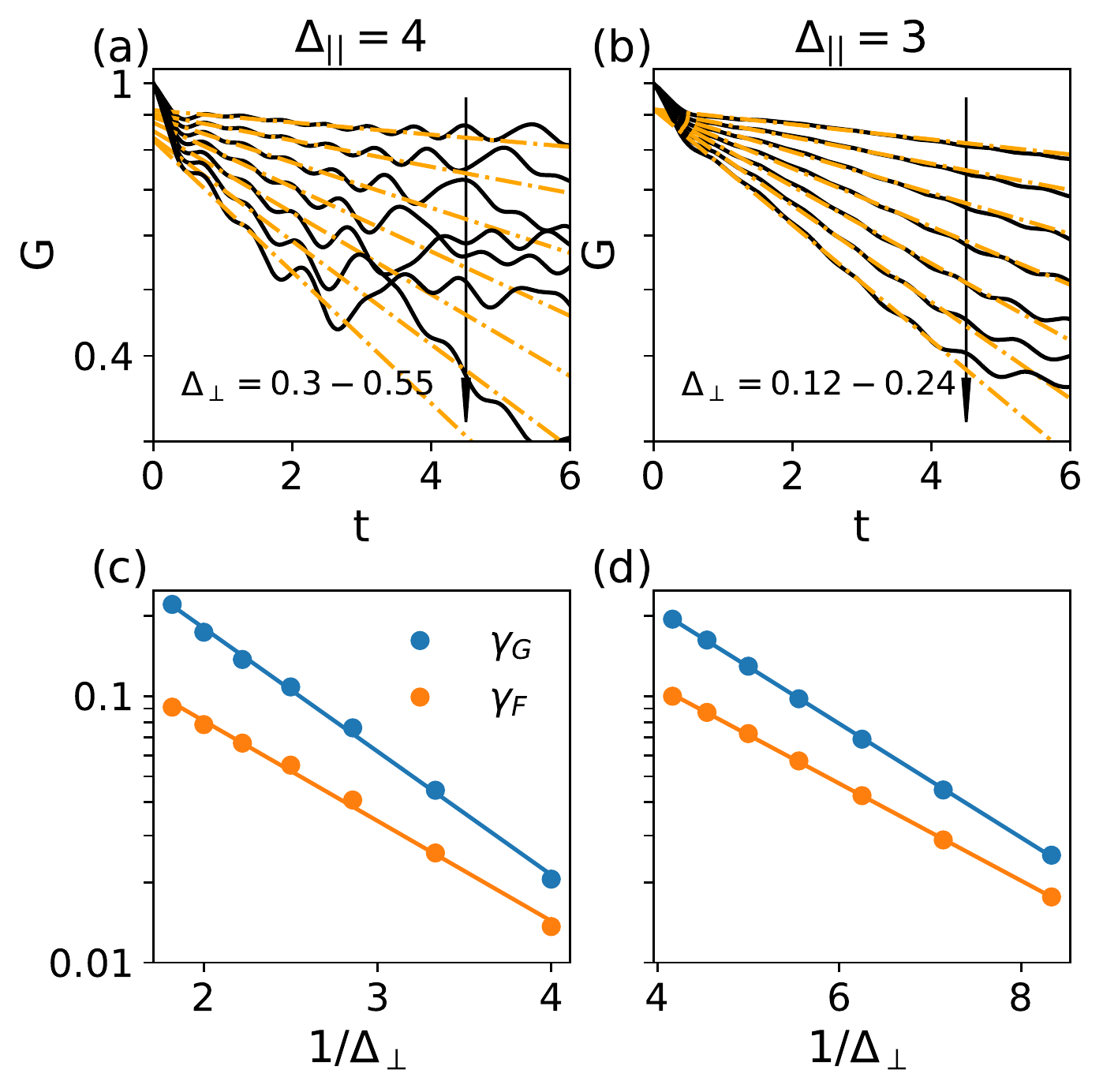}
    \caption{{\it False vacuum decay for the XXZ ladder.} {\it Panels (a) and (b):} Time evolution of  $G$ in Eq. \eqref{Gd} after a quench $\Delta_{\perp}\to-\Delta_{\perp}$ with $\Delta_{||}=4 $ (a) and $\Delta_{||}=3$ (b) with different values of $\Delta_{\perp}$. In (a) $\Delta_{\perp}=0.25,0.3,0.35,0.4,0.45,0.5,0.55$ while $\Delta_{\perp}=0.12,0.14,0.16,0.18,020,0.22,0.24$ in (b). The arrows indicate the growing direction of $\Delta_{\perp}$. 
    {\it Panels (c) and (d):} decay rates extracted from the fits in (a) and (b), respectively, on semi-logarithmic scale. The continuous lines are fits of the decay rates performed according to Eq.~\eqref{eq:td0}}
    \label{fig:ladder}
\end{figure}

In analogy with the Ising model, we prepare the false vacuum as the ground state at $-\Delta_{\perp}$ and then we quench $-\Delta_{\perp} \to \Delta_{\perp}$.
For several values of the interactions $\Delta_\perp$ and $\Delta_{||}$, 
we extract the decay rates $\gamma_{F,G}$ for $F(t),G(t)$ (here $F$ in Eq. \eqref{Fd} is built with the staggered magnetization 
and $G$ in Eq. \eqref{Gd} with the reduced density matrix of two adjacent rungs).
In Fig.~\ref{fig:ladder}, a) and b), we show the time evolution of $G$ after the quench for two values of $\Delta_{||}$. 
Even though we do not have analytic predictions for this ladder, we expect that the underlying mechanism of the false vacuum decay is the same so
we can fit the decay rate with Eq. \eqref{eq:td0} with the replacement $h_z\to \Delta_{\perp}$.
The test of this scaling for $\gamma_G$ is presented in Fig.~\ref{fig:ladder} c) and d), showing a perfect agreement. 
The quality of the fit for $\gamma_F$ is very similar, although in  Fig.~\ref{fig:ladder} we only report the final values for $\gamma_F$ and not the 
 data for $F(t)$.

\paragraph{Conclusions and outlook--- }
In this Letter we provided robust numerical evidence that for two one-dimensional spin models featuring confinement of elementary excitations it is possible to identify a range 
of physical parameters such that the rate of false vacuum decay is accessible in measurable time scales. 
The quench protocol that we described here is amenable to quantum simulation, for example with trapped ions or Rydberg atoms (both can simulate a system with confinement). 
For the false vacuum preparation, the  imaginary time evolution used in the numerics can be replaced by an adiabatic preparation.

We conclude by briefly discussing how the the trapped-ion quench experiment of Ref. \cite{tan2019} (for the observation of domain wall confinement in real time)
can be adjusted to measure the false vacuum decay.  
In this experiment, the ion dynamics is well captured by a long-range quantum Ising model in which the $\mathbb{Z}_2$ symmetry is spontaneously (and not explicitly) broken.
Hence, there are two degenerate real vacua and no false one. 
In order to get a phenomenology similar to our setup it is sufficient to slightly tilt the effective magnetic field (that in Ref. \cite{tan2019} is in the $z$ direction)  via a Rabi rotation, 
see the review~\cite{rev-ion}.
This tilting provides a small component of the magnetic field along the $x$ axis that breaks the degeneracy of the two vacua with a real and a false one.
Then the preparation of the system in the false vacuum and the following quench are done with the very same techniques 
exploited already in Ref. \cite{tan2019}.  Finally one- and two-point functions of the spin can be measured, as already done in Ref. \cite{tan2019}, giving access to
$F(t)$ and $G(t)$ in Eqs. \eqref{Fd} and \eqref{Gd}.

\begin{acknowledgments}
{\it Acknowledgments}. 
We thank Guido Pagano, Marcello Dalmonte, and G\'abor Tak\'acs for useful discussions. 
GL thanks BME and MK thanks SISSA for hospitality.
We thank ERC for partial support under grant number 758329, AGEnTh, (FS) and 771536, NEMO, (GL and PC).
MK acknowledges the Hungarian Quantum Technology National Excellence Program, project no. 2017-1.2.1- NKP- 2017-00001, and by the Fund TKP2020 IES (Grant No. BME-IE-NAT). MK acknowledges support by a Bolyai J\'anos grant of the HAS and by the \'UNKP-20-5 new National Excellence Program of the Ministry for Innovation and Technology 
from the source of the National Research, Development and Innovation Fund.

\end{acknowledgments}

\end{document}